\documentclass[a4paper,onecolumn,11pt,accepted=2021-11-12]{quantumarticle}
\pdfoutput=1
\usepackage{amsmath}
\usepackage{graphicx}
\usepackage[unicode=true]
 {hyperref}

\makeatletter
\usepackage[utf8]{inputenc}
\usepackage[english]{babel}
\usepackage[T1]{fontenc}

\usepackage{tikz}
\usepackage[numbers,sort&compress]{natbib}

\makeatother

\begin{document}
\global\long\def\id{\openone}%

\title{Opinion: Democratizing Spin Qubits}
\author{Charles Tahan}
\thanks{or, How to make semiconductor-based quantum computers without fabricating quantum dot qubits. Based on talks at the ARO/LPS Quantum Computing Program Review, August 2018, and 4th School and Conference on Spin-Based Quantum Information Processing, September 2018.}

\orcid{0000-0001-7726-8329}

\affiliation{Laboratory for Physical Sciences, 8050 Greenmead Rd, College Park,
MD 20740}
\email{charlie@tahan.com}

\maketitle
\selectlanguage{english}%
\begin{abstract}
I've been building Powerpoint-based quantum computers with electron
spins in silicon for 20 years. Unfortunately, real-life-based quantum
dot quantum computers are harder to implement. Materials, fabrication,
and control challenges still impede progress. The way to accelerate
discovery is to make and measure more qubits. Here I discuss separating
the qubit realization and testing circuitry from the materials science
and on-chip fabrication that will ultimately be necessary. This approach
should allow us, in the shorter term, to characterize wafers non-invasively
for their qubit-relevant properties, to make small qubit systems on
various different materials with little extra cost, and even to test
spin-qubit to superconducting cavity entanglement protocols where
the best possible cavity quality is preserved. Such a testbed can
advance the materials science of semiconductor quantum information
devices and enable small quantum computers. This article may also
be useful as a light and light-hearted introduction to quantum dot
spin qubits.
\end{abstract}

\section{Introduction}

\begin{figure*}
\includegraphics[scale=0.55]{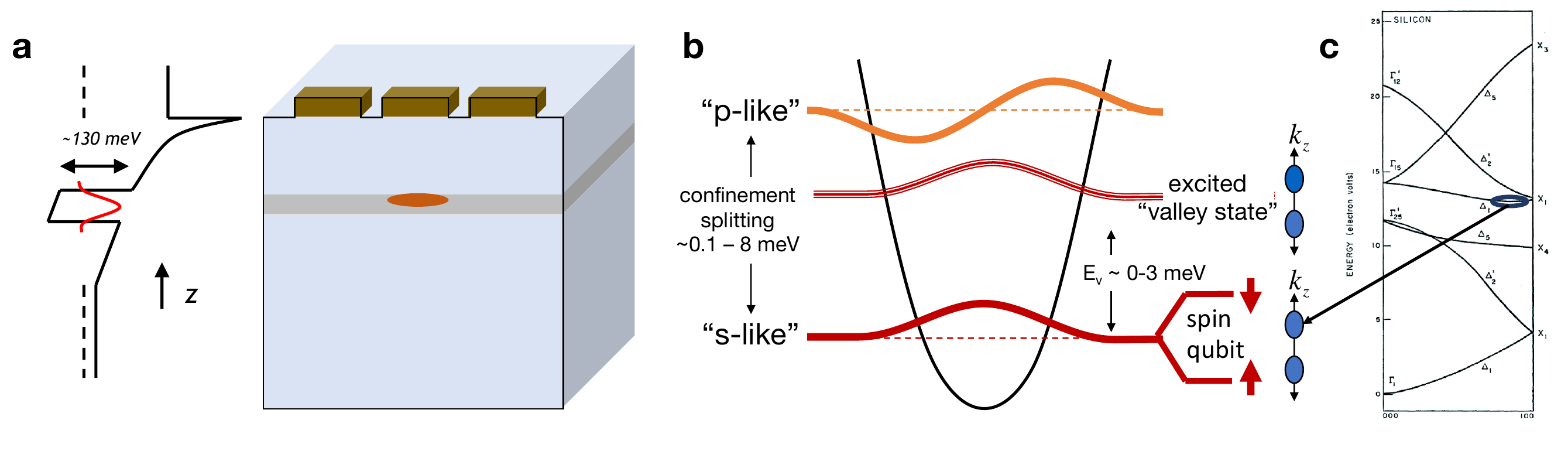}\caption{\textbf{Spin qubit in silicon. }a) A trapping potential due to a heterostructure
box or donor atom hosts our electron. For SiGe barriers above and
below a strained-Si quantum well, the trapping potential is roughly
50 meV per 10\% Ge in the barriers. b) The energy levels can be labeled
like an atom if we assume an effective mass theory \cite{BOOK-YuCardona}
(note that the real wave function of the electron also has Kohn-Luttinger
oscillations \cite{KL-PhysRev.98.915,PhysRevB.89.075302} due to the
conduction band valleys at \textasciitilde 0.8\textbf{$k_{max}$
}and atomistic oscillations from the crystal underneath this envelope).
The \textquotedblleft valley-splitting\textquotedblright{} between
the states can range from 0 to several meV. c) The conduction and
valence band symmetry governs possible additional levels. For electrons
in strained-silicon quantum wells or at inversion layers, there are
two occupied conduction band minima and thus double the number of
states (the other four states are pushed much higher in energy due
to the strain or confinement, respectfully).\label{fig:spin-qubit}}
\end{figure*}

The two states of a qubit are realized in the spin of an electron,
spin up and spin down. A single electron can be trapped in a semiconductor
box, called a quantum dot (Figure \ref{fig:spin-qubit}). In silicon---currently
the most promising material for spin-based quantum computing---the
indirect band-gap means that the electron has extra nearby energy
levels due to different combinations of the conduction band minima
or ``valleys'' where the electron exists. Temperature, noise, and
gate operations can cause unwanted excitation into these states. This
so-called ``valley splitting'' problem, especially in silicon-germanium
quantum dots, impacts yield, initialization/readout, and quantum operations,
and originally motivated this work. Although obscure, this materials
science issue is a roadblock to quantum information processors in
silicon, and valley splitting is only representative of a greater
challenge.

Like for other qubit parameters such as coherence time and operation
error rates, to measure the valley splitting one must fabricate a
quantum dot and test it, typically at dilution refrigerator temperatures\footnote{Around 50 to 100 milli-Kelvin or about 60 times colder than deep space.}.
The general difficulty in making qubits in semiconductors has hampered
progress in the field. Exciting recent successes---functional dot
qubits and compelling quantum gate demonstrations \cite{HansonSpinsfewelectronquantum2007,BorselliUndopedaccumulationmodeSi2014,Zajacreconfigurablegatearchitecture2015,VeldhorstTwoQubitLogic2015,ReedReducedSensitivityCharge2016,SOPKuemmeth-PhysRevLett.116.116801,RudolphCouplingMOSQuantum2016,ZajacQuantumCNOTGate2017,Watsonprogrammabletwoqubitquantum2017,zwanenburg-siqe-2013,2013-kloeffel-loss,2017-yoneda-tarucha}---have
taught us a lot about how to make good qubits. Yet there is a high
barrier to entry for new experimental groups compared to, say, superconducting
qubits.

\begin{figure}
\centering{}\includegraphics[scale=0.4]{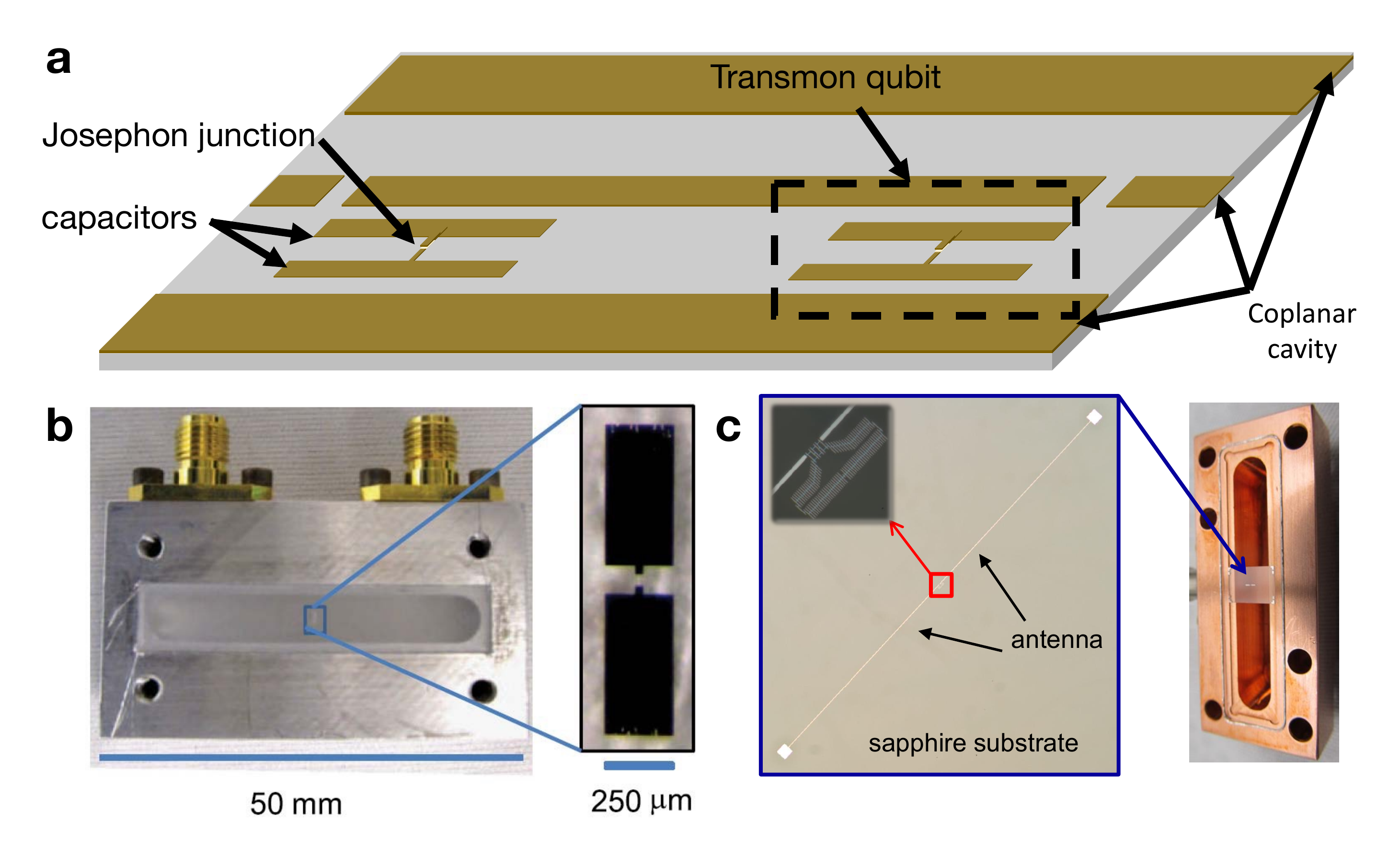}\caption{\label{fig:transmon-wireless}\textbf{From the transmon qubit to the
wireless revolution. }(a) The transmon superconducting qubit is a
single layer of metal and a shadow-evaporated junction. It is extremely
robust in that nearly all fabricated qubits work. In a co-planar geometry
a superconducting cavity can be used to readout the qubit (circuit
QED). (\textbf{b}) A crucial step to advance the field was applying
these same ideas to a 3D geometry. By going larger the participation
of loss mechanisms was proportionally lowered and allowed systematic
study of qubits with longer life while the more scalable 2D qubits
were catching up in performance. But perhaps more interesting, the
wireless approach can allow different types of qubits to be tested
more quickly, even as in (c) fluxonium, where there isn't a natural
coupling to an electric field (an antenna can be used to couple to
the 3D field). (Figures 2b and 2c courtesy Yale.)}
\end{figure}

\begin{figure*}[t]
\centering{}\includegraphics[scale=0.45]{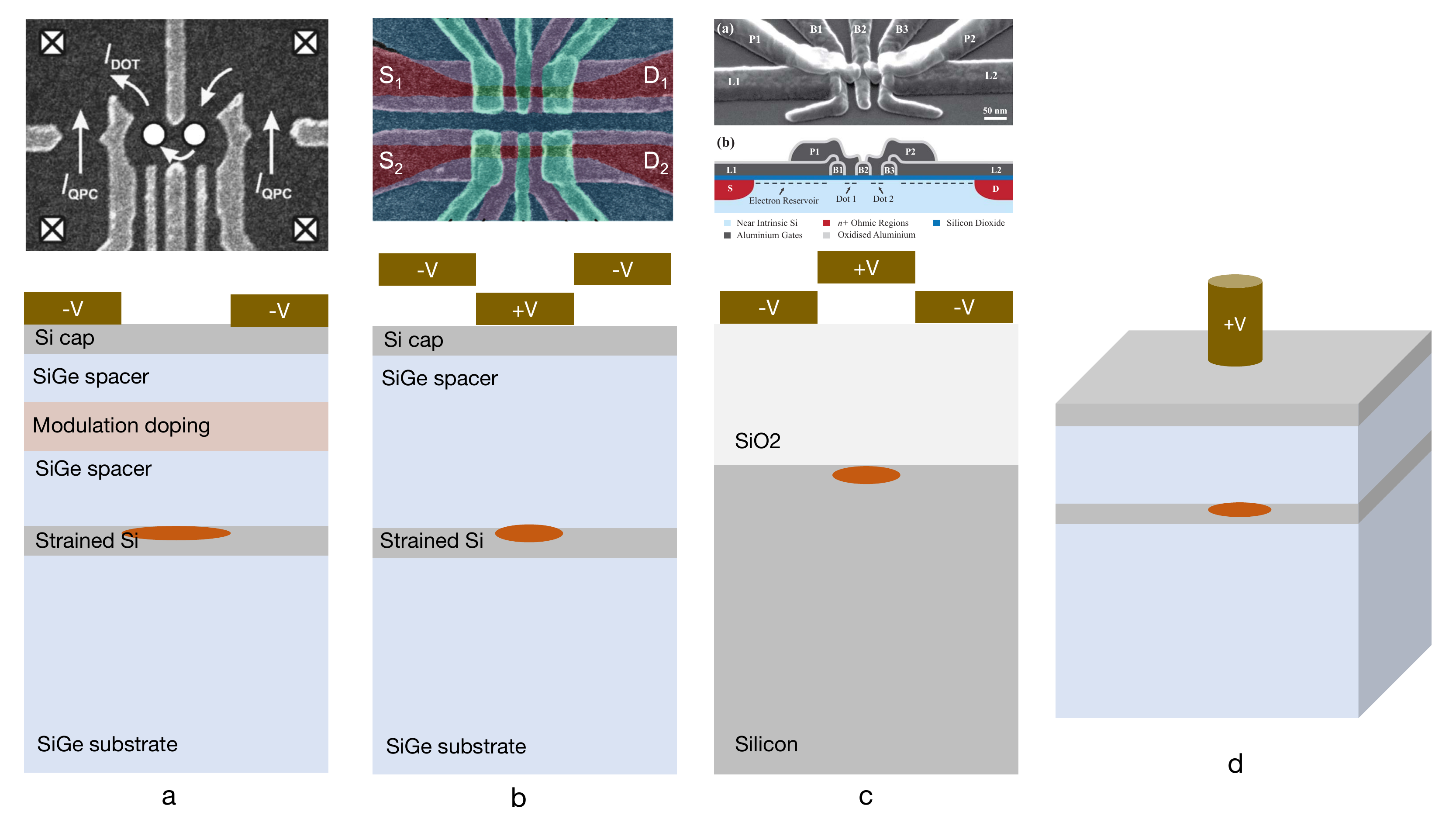}\caption{\label{fig:QD-simplified}\textbf{Quantum dot, simplified.} a) Early
depletion mode quantum dots (modeled on GaAs quantum experiments before
them) utilized doping layers (modulation doping) and negatively charged
metal top gates to push away all electrons but one \cite{FriesenPracticaldesignsimulation2003b};
these suffered from severe disorder when implemented in silicon. b)
Accumulation mode devices use overlapping gates to form dot and tunnel
gates and represent the current state of the art \cite{LaiPauliSpinBlockade2011,BorselliUndopedaccumulationmodeSi2014,Zajacreconfigurablegatearchitecture2015};
often a nearby charge sensor (also a dot) is used for spin-charge
readout. Here, the doping layer can be removed if implanted electron
sources (donor-implanted regions in the crystal) are nearby. c) A
single quantum dot potential in it's simplest form. d) A quantum dot
created with a single gate wire whose electrostatic potential has
already trapped an electron in the quantum well. (Top row device images
are courtesy TUDelft, Princeton, and UNSW.)}
\end{figure*}

Do superconducting qubit experimentalists have it easy? Yeah, they
kind of do. The robust\footnote{That is, if you make a transmon qubit, they typically work.}
transmon qubit \cite{KochChargeinsensitivequbit2007,SchreierSuppressingChargeNoise2008,HouckLifechargenoise2009}
requires only one layer of metal, with large lithographic dimensions
\cite{PaikObservationhighcoherence2011}. The transmon and variants
can be characterized and controlled with a single microwave cavity/generator/line
or even wirelessly via a properly designed superconducting cavity
(Figure \ref{fig:transmon-wireless}). Superconducting circuits can
be floating, requiring no source or sink of carriers. In contrast,
consider what\textbf{ }we know now as best practices for making a
lateral silicon quantum dot qubit (Figure \ref{fig:QD-simplified}).
One needs to make very small dots (due to the relatively large effective
mass of silicon), with multiple layers of overlapping or tightly-aligned
metal gates to limit cross-capacitance between dot gates\footnote{Note that ``gates'' is used in two ways. First, there are quantum
gates, also called operations, that change the state of the qubit.
The second way refers to the physical metal wires that (by changing
voltage potentials) trap or move the electron in the channel underneath.}, all with O(10 nm) wire-widths at 50-100 nm wire pitch. Poor yield
to disorder is a major problem.\textbf{ }Worse, the materials stack
is critically important to whether the dot qubits work at all or have
desirable qubit properties. Spin-based quantum dot qubits also need
multiple physical wires per dot and often nearby charge sensors (for
spin-to-charge conversion-based readout \cite{PettaCoherentManipulationCoupled2005}).
This level of complexity in fabrication---which \emph{must} be coupled
with good materials science properties of the wafer and the gate stack---retards
both new qubit exploration and characterizing many, individual quantum
dots to optimize materials parameters.

We can separate the materials science challenge from the ``making
the dots and measure them'' challenge. If the dots and associated
readout circuitry can be made on another circuit chip or board, then
the actual qubit-hosting wafer can be optimized separately. (And even
made of different crystals such as germanium). The idea of ``flip-chip''
engineering has already been applied successsfully in the superconducting
context \cite{Rosenberg3Dintegratedsuperconducting2017,FoxenQubitcompatiblesuperconducting2017},
while the concept of a ``probe'' trap has been used in ion trap
quantum computing \cite{ArringtonMicrofabricatedstylusion2013,HiteMeasurementstrappedionheating2017}
to search for heating mechanisms on relevant surfaces. In semiconductors,
Ref. \cite{eng_high_2005} and later Ref. \cite{2016-kouwenhoven-noninvasive}
explore non-invasive techniques for gating pristine materials.\textbf{
}Here, we motivate further a testbed where dots are \emph{induced
and measured} by a separate chip to characterize materials \cite{article-shim-induced-dot-APL}
and qubit approaches. In doing so, an acceleration of progress similar
to that driven by the ``wireless'' 3D transmon (Figure \ref{fig:transmon-wireless}b)
could be replicated in the spin community.

There are other ways to democratize spin qubits. One can distribute
known good devices to academic groups interested in exploring new
qubit encodings and ways to control them; this is the foundry model\footnote{Laboratory for Physical Sciences / Army Research Office REQUEST FOR
INFORMATION \href{https://sam.gov/opp/98673a00eabb4ffda553f133e5211d77/view?keywords=\%22Qubits\%20for\%20Computing\%20Foundry}{Qubits for Computing Foundry}
(QCF), 2020}. Increasing throughput of testing at multiple temperature stages
(room temperature, 1 to 4K depending on the physics, and \textless 100
mK) also directly benefits fabricated device optimization. Recent
work toward a cryo-prober \cite{2020-cryoproberIntel}, particularly
at 1-2K where dot physics starts to correlate with qubit performance,
will accelerate yield of critical parameters. However, dilution refrigerators
(DR) are needed for the most advanced qubit measurements (relevant
to target applications). Cryogenic switches inside the DR will allow
for a handful of quantum dot systems to be measured in sequence without
a fridge warm-up. 

\section{Making a quantum dot qubit without fabricating a quantum dot}

A spin qubit is formed from the Zeeman split sub-levels (see Figure
\ref{fig:spin-qubit}b) of the ground state of an electron trapped
in some potential inside a semiconductor. That electrostatic potential
can be artificial, formed from the combination of a heterostructure\footnote{A stack of differing crystals or crystal alloys.}
and an external voltage \cite{LossQuantumcomputationquantum1998a},
or natural, the pull of an implanted donor \cite{Kanesiliconbasednuclearspin1998a}.
The lowest levels of one or more electrons or holes inside a single
dot can form a qubit. Alternatively, a qubit can be ``encoded''\footnote{The notion of encoded or ``decoherence-free subspace or subsystem''
(DFS) qubits is not exclusive to spins. DFS encoding provides some
immunity to global fields or errors, although there is significant
experimental evidence that the dominant noise sources affecting dot
qubits currently are local to each dot. The main benefits for spins
are: 1) spins are small, so microwave gates can result in a lot of
cross talk, 2) microwave gates (driving single dot single spin rotations
with a magnetic field) are slow, 3) two qubit gates via the exchange
interaction are fast. Terminology-wise I prefer 2-DFS, 3-DFS, etc.
as denoting encoded qubits where the dots are isolated from each other
when not undergoing (pair-wise only) encoded qubit operations. Other
terminology is often used such as singlet-triplet qubits (for double
dots) and exchange-only (for triple dots). More recently the community
has been revisiting using small clusters of spins which form molecules
with always-on exchange couplings between dots starting with the resonant
exchange (3 dot) qubit (single sweet spot) and derivatives like the
AEON qubit (double sweet spot). Because the exchange interaction is
always on, these qubits are more sensitive to charge noise in their
off state than an isolated spin, always, but have other potential
advantages.} into the larger Hilbert space of multiple separate quantum dots (e.g.,
\cite{Lidar-PhysRevLett.81.2594,DiVincenzo:2000uz,Levy-PhysRevLett.89.147902,FongUniversalQuantumComputation2011,TaylorElectricallyprotectedresonantexchange2013,ShimChargenoiseinsensitivegateoperations2016,RussThreeelectronspinqubits2017}).
Each approach has different, potentially useful properties for quantum
computing. Typically, 1 or 3 electrons are loaded per dot\footnote{Increasing the (odd) number of electrons keeps the overall spin 1/2
but also lower the excited energy levels which can cause complications
in gates.} in a roughly parabolic potential in the plane of the two-dimensional
electron gas (2DEG) or inversion layer. The potential must be deep
enough such that the excited state orbital levels (``p-like'' envelope
functions\footnote{Effective mass theory lets us throw many of the complexities of actually
being in a crystal under the rug and bundle that up in a new mass
with associated hydrogenic-like orbital levels.}) are well above the energy of thermal excitations ($\sim kT$). In
an indirect band-gap semiconductor like silicon, the electron wave
function is a superposition of all the conduction band minima (or
valleys). This valley degeneracy is lifted by strain or an abrupt
potential. (Alternatively, holes exist at the top of the valence band
at the $\Gamma$-point and can be considered spin-$3/2$ particles,
although there are sometimes nearby heavy-hole or SOC split bands.)
Somewhat surprisingly, the spin splitting (linearly proportional to
the magnetic field) can be \emph{smaller} than the effective temperature
and typically is in many experiments\textbf{ }\cite{HansonSpinsfewelectronquantum2007}.
Kramers degeneracy (where a ``forbidden'' very tiny matrix element
connects phonons between the two spin states) results in extremely
long lifetimes for spins in some solids, a fact that has been known
at least since the 1950s \cite{FeherElectronSpinResonance1959a,THESIS-Tahan}.
(The same can also be true for some hole states, e.g. \cite{ruskov_-chip_2013}.)
\begin{figure*}
\centering{}\includegraphics[scale=0.44]{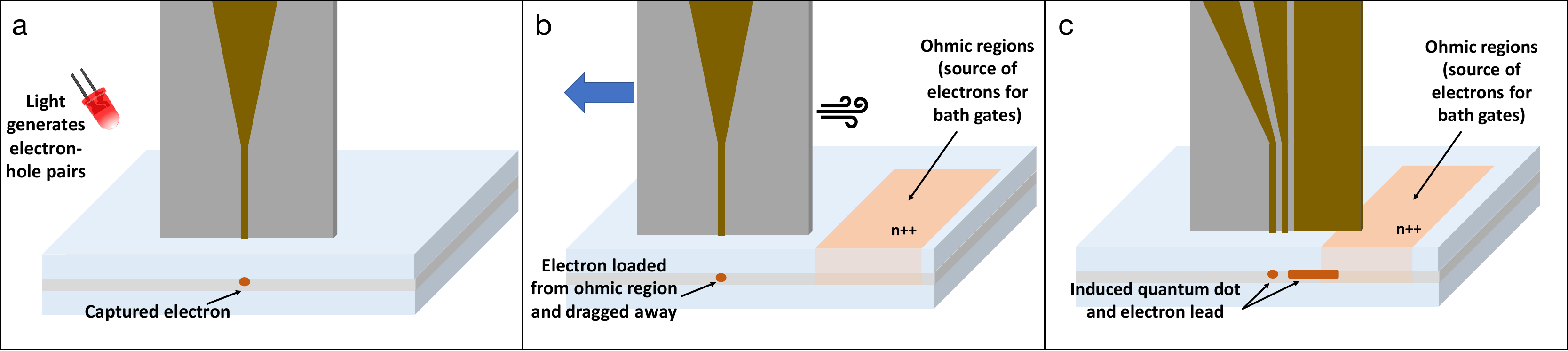}\caption{\label{fig:loading-an-electron}\textbf{Loading an electron into an
induced quantum dot from a perpendicular gate chip.} (a) Light above
the band-gap can put electrons into the conduction band (or background
dopants can create a reservoir to load from). (b) Implanted regions
on the chip can be used as loading zones if the tip/gate-chip has
the ability to move. (c) By adding a bath gate to the gate-chip, one
can load electrons even from a long distance away.}
\end{figure*}

When designing a quantum computer based on quantum dots, we care about
a number of intrinsic parameters derived from the quantum dot potential,
material and material stack, the proximity of other qubits/defects/oxides/gates,
and other parameters that are relevant for two qubit operations. These
parameters include the level structure of the dot (or spectroscopy
of the excited states), the spontaneous decay time (called the $T_{1}$
time, almost always due to emission of a phonon at these dilution
refrigerator temperatures, but more complicated above \textasciitilde 1K),
and the decay of the coherence of the qubit as a function of time
(or $T_{2}^{*}$ for the specific 1/e time assuming an exponential
fall-off, which is actually not usually the case). All these parameters
can change with the number of particles per dot and if multiple dots
are coupled together, either via the Pauli exchange interaction or
capacitively. 

Let us consider the simplest quantum dot, using best practices from
recent progress (Figure 3). Our options include semiconductors such
as GaAs, silicon, germanium. GaAs has spinful nuclei, leading to poor
$T_{2}^{*}$ times (\textasciitilde ns). Germanium is interesting
as discussed below. But the pull of silicon is strong due to the CMOS
industry (ultra-chemically pure and perfect crystals, precision lithography,
good dielectrics) and the fact that isotopically enriched silicon
exists and is available (where the spin-1/2 Si$^{29}$ nuclei have
been removed leaving only spin-0 silicon-28), leading to extremely
long coherence times \cite{Tyryshkin:2012aa}. 

First, create an electrostatic trap for the electron in the growth
direction, $z$, by making either a ``quantum well'' or an inversion
layer. For the former, use a strain-engineered SiGe-Si-SiGe sandwich
\cite{SchafflerHighmobilitySiGe1997}. For the latter, use the well-known
oxide-silicon (MOS) interface of silicon and silicon-dioxide. An accumulation
gate (positive voltage) creates a triangular-like potential, pulling
electrons (if available) up the top against the trapping interface
(Figure \ref{fig:spin-qubit}a). There are no electrons until we put
them there (assuming no doping). Due to the effective mass of electrons
in silicon, to get an orbital splitting of \textasciitilde 1meV =
10K, we need an effective ``box'' of \textasciitilde 30 nm laterally.
Finally, a combination of negative (depletion) and positive (accumulation)
gates can form the trap in the $(x,y)$ plane creating our dot potential.
Because of the large effective mass of electrons in silicon compared
to GaAs\footnote{GaAs was the physicists semiconductor and III-V heterostructures the
host for the quantum hall effect, the fractional quantum Hall effect
(e.g., Noble prizes) and the first artificial atoms or quantum dots.}, the community has learned that unwanted ``dots'' (think a disordered
eggshell container) can form in the quantum well via the presence
of donors, or by possibly tiny strain due to metallic gates \cite{ThorbeckFormationstraininducedquantum2015}
on the surface. Therefore, don't dope the quantum well. Instead, implant
donors in source regions creating a bath of electrons that can be
brought electrostatically near the active quantum dots with gates.
And make the gates as uniform (``total coverage'') and as far away
as possible. This is how the best arrays of lateral quantum dots for
qubits are fabricated today (Figure \ref{fig:QD-simplified}).
\begin{figure}
\includegraphics[scale=0.5]{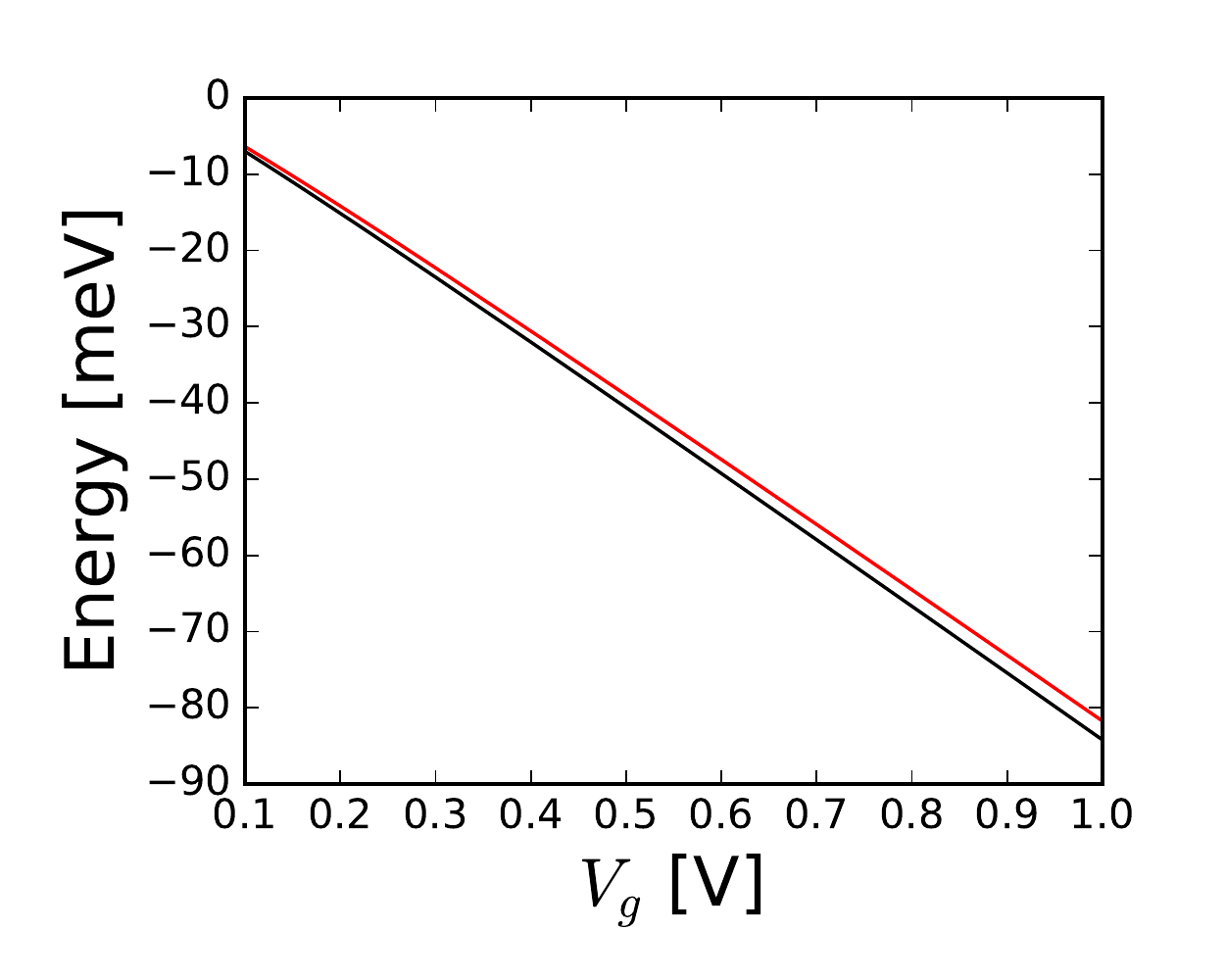}\includegraphics[scale=0.26]{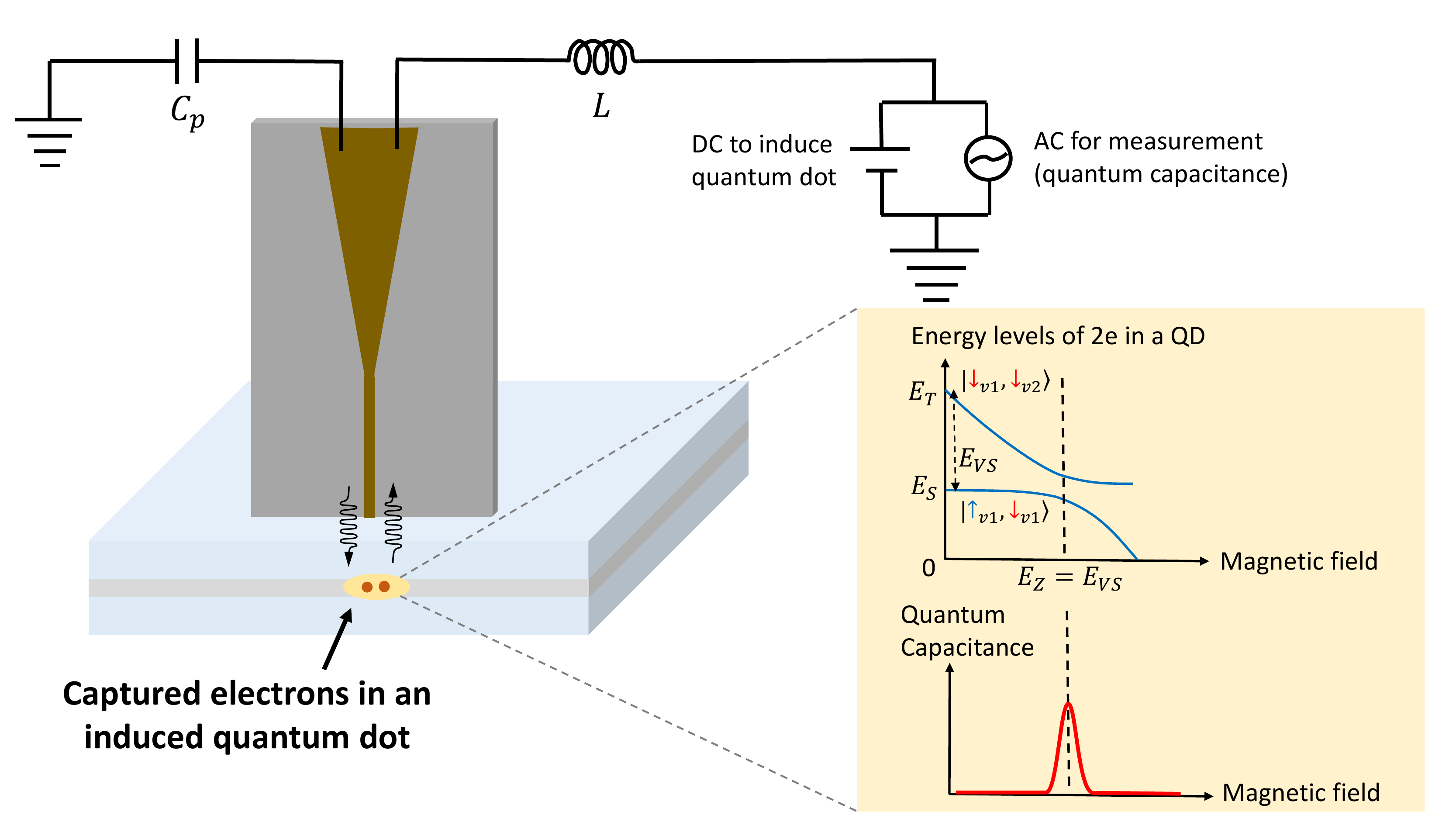}

\caption{\label{fig:readout} \textbf{Using quantum capacitance for dot characterization.
}(\textbf{left}) Simulated energy bands of lowest two states versus
gate voltage of induced single electron quantum dot of Figure 3d.
Although the curvature is very small, it exists and may be measurable
\cite{article-shim-induced-dot-APL}. The black line is the curvature
of the ground state and the red line is the curvature of the first
excited state. (\textbf{right}) Energy curvature as a function of
magnetic field near the valley splitting and Zeeman splitting anti-crossing
(when they are equal) in a quantum dot with two electrons, where a
large quantum capacitance signal will be observed. The curvature vs.
probe gate voltage and the curvature vs. magnetic field are proportional
to each other, so a large quantum capacitance signal will also be
seen through the gate voltage in the reflected signal. (Figures courtesy
Rusko Ruskov and Yun-Pil Shim.)}
\end{figure}

To \emph{induce} a dot potential, only a single wire is needed (Figure
\ref{fig:QD-simplified}d). This can be formed from a metal gate on
the dot wafer surface, \emph{or,} by a wire on a different chip or
probe tip (with a positive voltage, appropriate geometry, and distance
from the target active layer \cite{article-shim-induced-dot-APL}).
Indeed, Scanning Tunneling Microscope (STM) tips have been used in
the past to induce dots on the surface of materials \cite{PhysRevB.59.8043,SalfiValleyfilteringspatial2017a}.
Thus, the simplest dot doesn't require fabricating a dot on the ``dot
chip'' at all. Instead, fabricate a ``dot inducing chip'' or \emph{gate
chip} on a separate substrate, with a single layer of metal. Then
place that perpendicular to the ``wafer chip'' (touching or not)
to where the dot is actually intended to be created. All the circuitry
we need to measure the dot can be off the wafer chip under test.

Consider three (increasingly demanding) options for loading electrons
into the induced quantum dot (Figure \ref{fig:loading-an-electron}).
Option 1: shine light with energy above the band-gap on the wafer
to generate carriers\footnote{Light is often used in quantum dot experiments to ensure that the
well is populated or to increase density in the 2DEG and is also common
practice in STM experiments. Light background doping (not recommended)
or a doped back gate reservoir are also possible.}, Figure \ref{fig:loading-an-electron}a. Generate enough carriers
and our dot will trap one (or more, depending on the depth of it's
trapping potential). Option 2 (perhaps only suitable to STM-like,
single tip, set-ups): Figure \ref{fig:loading-an-electron}b, move
the inducing gate to an implanted region on the wafer and physically
move the loaded dot away from it to isolate the dot. Option three:
add other lead(s) to the gate-chip, including a much wider and fatter
``bath'' gate (Figure \ref{fig:loading-an-electron}c) that can
bring electrons into the channel from the implanted region to the
dot much like is already done. Implanting is a standard procedure
and requires a mask but can be outsourced. (So we've broken the non-invasive
pledge for options two and three.) 
\begin{figure}[h]
\centering{}\includegraphics[scale=0.45]{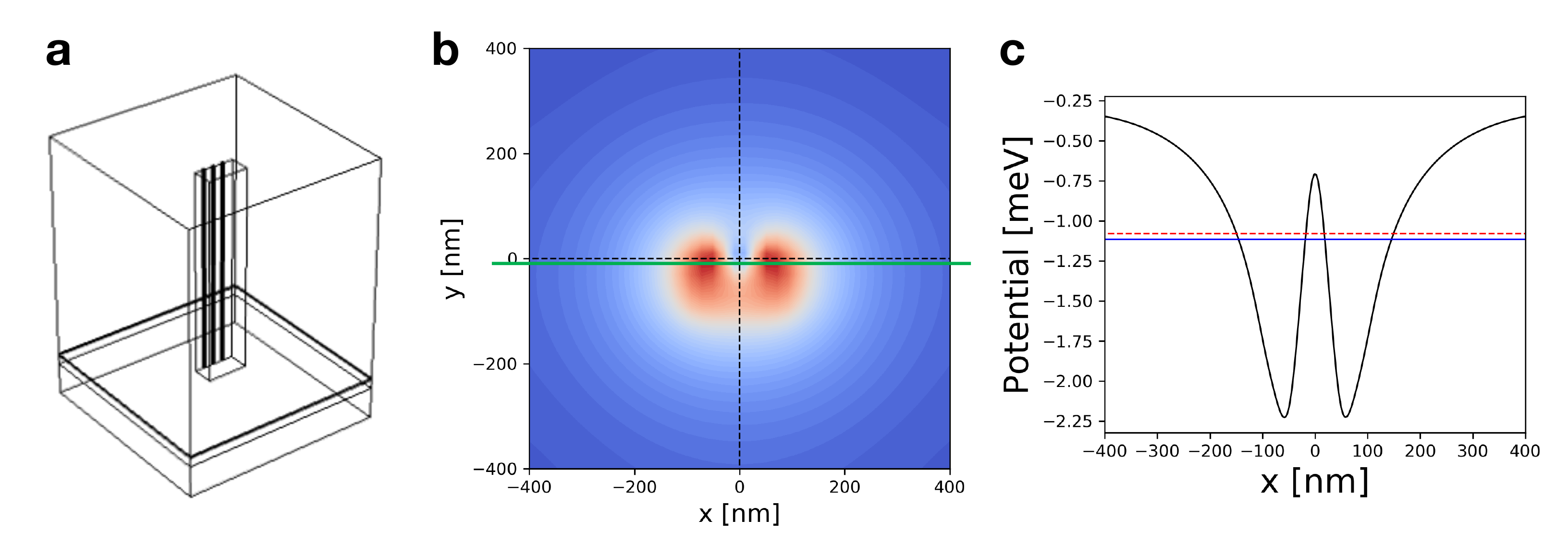}\caption{\label{fig:2dotchip} \textbf{2 quantum dot simulation. }(a) Gate-chip
is at 10 nm above the heterostructure, each wire has 10 nm by 10 nm
cross-section. Wire-wire distance is 50 nm. Applied voltages on the
3 wires are 0.02 V, -0.02 V, 0.02 V. b) Electrostatic potential of
the induced double dot system in the $xy$ plane of the wafer two-dimensional
electron gas (2DEG) in a silicon-germanium quantum well. c) Electrostatic
potential along the green line. The solid blue line is the ground
state energy and the red dashed line is the first excited state energy.
(Figure courtesy Yun-Pil Shim.)}
\end{figure}

\section{Characterizing an induced quantum dot}

An electron must be present in the induced quantum dot to have a qubit.
If the dot is truly isolated with only one wire to both maintain the
potential and probe the dot, that is really hard to do. Best practice
for qubit measurement is to create a nearby quantum dot charge sensor
(see the top-down SEM image in Figure \ref{fig:QD-simplified}b for
an example).\footnote{The dot is tuned to the edge of a Coulomb blockade peak for maximum
sensitivity (such that small charge redistributions in the qubit dots
affect the current flowing through the dot sensor). } By pulsing the electrons in the dots one can figure out how to convert
the spin information to different charge distributions, detectable
with the readout dot. However, since we only have a single wire, a
better option is to use so-called (in the dot community) dispersive
readout. A tank circuit is attached to a nearby dot gate. Small changes
in the quantum capacitance as seen by that gate (or more generally
the curvature coupling \cite{RuskovQuantumlimitedmeasurementspin2017})
are detected by measuring the phase shift of a reflected rf-pulse
(dispersive shift of the resonator) at the resonator frequency (where
the frequency is relatively low, 100 Mhz to 500 MHz) \cite{ReillyFastsinglechargesensing,PeterssonChargeSpinState,CollessDispersiveReadoutFewElectron2013,Gonzalez-ZalbaProbinglimitsgatebased2015,Gonzalez-ZalbaGateSensingCoherentCharge2016,RossiDispersivereadoutsilicon2017,MizutaQuantumtunnelingcapacitance2017,SchaalConditionaldispersivereadout2017}.
Note that this technique as applied so far has only worked in two
ways: detecting a signal when electrons tunnel in and out of a nearby
bath (so-called ``tunneling'' capacitance) or at a charge transition
(e.g. between dot occupations $1,1$ to $2,0$, see Figure \ref{fig:multidot_exp}
for a preview) to a nearby dot.\footnote{where the quantum capacitance is detected, the curvature of the two-dot,
one-electron system is maximal at the degeneracy point, the symmetric
state of a charge qubit.} These signals map out 0, 1, etc. electrons in the dot. 

A higher-frequency resonator can also be used as in \cite{MiStrongCouplingSingle2017,Samkharadze1123},
putting the cavity into the quantum regime (and increasing compatibility
with superconducting amplifiers). Ref \cite{article-shim-induced-dot-APL}
showed that it may be possible to measure the small energy band curvature
(Figure \ref{fig:readout}a) due to a single, stationary electron
as compared to no signal (no electron trapped). Assuming the use of
a superconducting resonator in series with the dot-inducing gate,
quantum capacitances as low as 0.01 attoFarad should be observable
if a $Q\sim10^{5}$ can be achieved (see \cite{article-shim-induced-dot-APL}
for details), allowing for the detection of electrons with a single
lead without tunneling transitions to a reservoir or dot (which has
never been seen before). If the measurement apparatus is fast enough
then the appearance of an electron from the bulk (due to light-created
carriers for example) may be observable. If these don\textquoteright t
work, then using the bath gate or a nearby dot (a 2 dot probe) will
be necessary, more complicated, but making all the known techniques
for qubit characterization available. 
\begin{figure}[t]
\noindent \centering{}\includegraphics[scale=0.4]{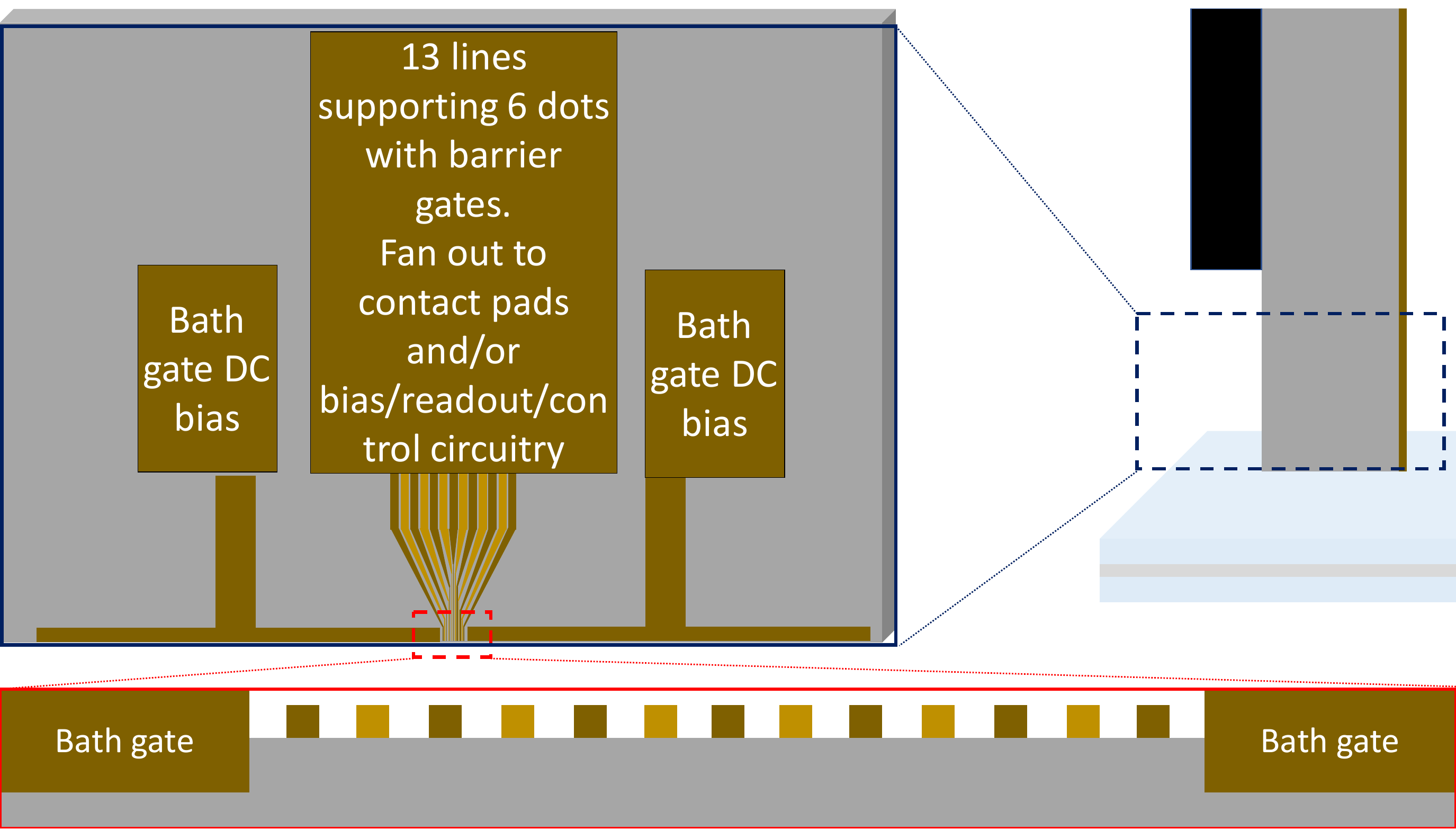}\caption{\label{fig:6dotchip} \textbf{6 quantum dot \textquotedblleft testbed\textquotedblright .
}On a single chip wires to support 6 dots with barrier gates in between,
with bath gates, can be placed in a single layer of metal, along with
associated electronics. With six dots you can create two 3-dot encoded
qubits and try out many one and two-qubit encoded gate protocols.
(top) Side views of the probe chip. (bottom) View (cross section)
of the gate chip from the wafer under test.}
\end{figure}

With the electron number in the dot known, let\textquoteright s focus
on spectroscopy, or charting of the excited states of the dot. Doing
so solves our original problem of measuring the valley state across
a chip non-invasively. Ref \cite{article-shim-induced-dot-APL} proposes
solutions for a single wire. At the magnetic field that equals the
valley splitting energy, there is an anti-crossing which results in
a quantum capacitance change. With two electrons in the dot, the curvature
is even larger and occurs in the ground state (Figure \ref{fig:readout}b).
Detecting this curvature allows the valley splitting to be measured
by sweeping the magnetic field (or valley splitting via $E_{z}$ over
a smaller range). Doing precise spectroscopy requires a relative energy
scale. The magnetic field provides that if it can be well calibrated.
Another option is to introduce another rf field which drives transitions
in the dot. Unfortunately, given the large possible range of valley
splittings, and the possibility of very large orbital splittings (up
to 8 meV), tunability of the microwave field would have to be over
a vast range for a single dot. With two dots a new energy scale emerges,
the detuning between the two dots, and there are many more options;
valley spectroscopy can be achieved without a magnetic field \cite{BurkardDispersivereadoutvalley2016}.
The reflectometry approach allows one to measure the critical parameters
of the system even if the excited states are much larger than the
cavity frequency and for arbitrarily high valley splittings in a two
dot system.

To summarize, with a single wire we can detect if there is an electron
in the dot and measure the valley splitting given some rather stringent
requirements. There are other ways to measure valley splitting: already
demonstrated techniques like photon-assisted tunneling, that involve
multiple dots or tunneling to leads. Those can also be realized in
multi-lead systems.

\section{Characterizing an induced quantum dot quantum computer}

Imagine a one-dimensional array of dots induced by a gate-chip on
a chip wafer. Illustrations of such systems are shown in Figures \ref{fig:2dotchip}
and \ref{fig:6dotchip}. It is possible to make two dots with just
two wires, and for many years the community relied on detuning of
energy levels between dots to produce two-electron interactions. Now
it is understood that using a barrier gate to control electron wave
function overlap is better.\footnote{Loss and DiVincenzo \cite{LossQuantumcomputationquantum1998a} originally
proposed to use the barrier gate, although there wasn't an appreciation
of sweet spots at the time. The community took the easier path experimentally
and chose detuning which has several downsides.} Using the barrier gate allows for operation at a sweet spot (symmetric
operating point) \cite{ReedReducedSensitivityCharge2016,SOPKuemmeth-PhysRevLett.116.116801}
and it is less sensitive to charge noise than the plunger gates above
the dots \cite{ShimDetuningPhysRevB.97.155402} (the exchange interaction
is less sensitive to the tunnel barrier than to the detuning, so charge
noise is minimized by using only high speed lines on the barrier if
possible). Moving to multiple wires allows one to measure dot qubit
quantum properties such as coherence times and even do quantum operations,
that is, make a small quantum computer.
\begin{figure}
\begin{centering}
\includegraphics[scale=0.44]{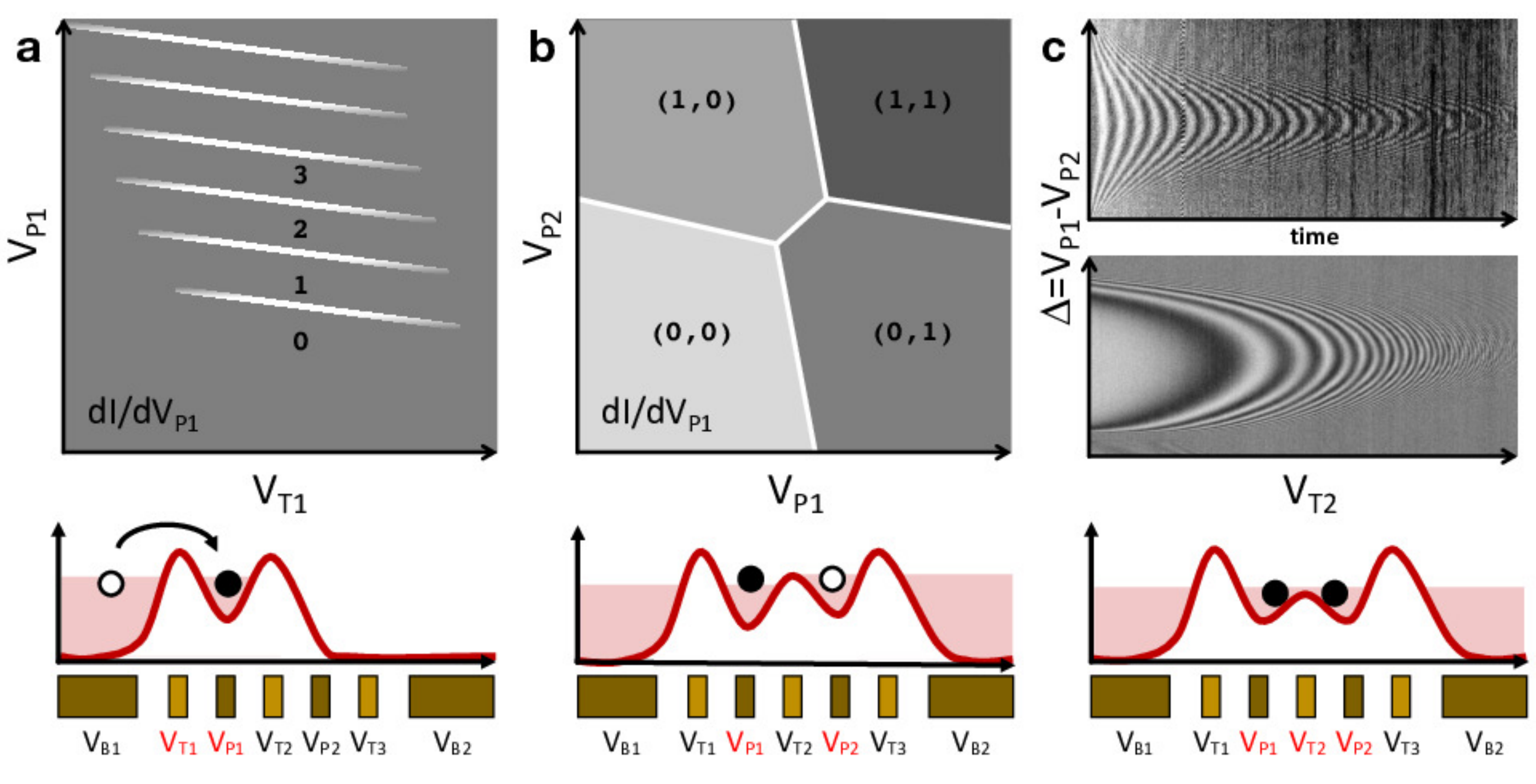}
\par\end{centering}
\caption{\label{fig:multidot_exp}\textbf{Multi-dot experiments that characterize
the system }(a) Loading electrons into a single dot by changing the
plunger gate above the dot versus the tunnel gate from the electron
bath, also called the charge stability diagram of a dot (P vs. T plot).
(b) Charge stability diagram of a double dot system for a given tunnel
barrier choice (P vs. P plot). (c) Upper plot shows coherent exchange
oscillations for a given tunnel barrier choice as a function of detuning
between the dots. At the symmetric operating point (SOP) or zero detuning
of the two dot potentials the number of coherent oscillations is enhanced
\cite{ReedReducedSensitivityCharge2016,SOPKuemmeth-PhysRevLett.116.116801}.\textbf{
}The lower plot shows a so-called \textquotedblleft fingerprint\textquotedblright{}
plot demonstrating the dependence of exchange on $\Delta$ and $V_{T2}$.
In this plot the average singlet probability is shown after evolving
for 500 ns at a potential specified by the axes \cite{ReedReducedSensitivityCharge2016}
(both c plots courtesy HRL).}
\end{figure}

Let's quickly go through a sequence of experiments that would characterize
the wafer in question (Figure \ref{fig:multidot_exp}). With a multi-wire
device it is natural and preferable to use a bath gate to load electrons.
In this context our first experiment is to chart out the charge stability
diagram for loading electrons from zero. To do this we vary $V_{P1}$
versus $V_{T1}$, where $V_{P1}$ is the dot gate and $V_{T1}$ is
a tunnel gate between the bath and the dot. One should see lines in
this plot indicating the transitions between $n-1$ and $n$ electrons
(Figure \ref{fig:multidot_exp}a). (The straighter the line the better,
it indicates small cross capacitance of the gates. Curviness means
changing cross capacitance which is very bad because the electron
is moving. It also makes it harder to dynamically compensate for such
gates \cite{Baart:2016aa,Mills:2019aa}.) Once we can reliably load
single electrons, then charge stability diagrams (P vs. P, see Figure
\ref{fig:multidot_exp}b) would be performed to map out the parameter
space of the two dot system. An increasingly useful metric once this
stage is reached is the charge-noise spectrum of each dot (e.g., \cite{Connors_PhysRevB.100.165305}),
which can be extracted via resonator measurements \cite{Mi_PhysRevB.98.161404}.

To do quantum coherent measurements we need to actually measure the
spin qubits. Incorporating readout allows one to experimentally determine
the coherence time ($T_{2}$) and lifetime ($T_{1}$) of the qubits
as well as the error rates of one and two-qubit operations with the
right pulse sequence. Combining quantum capacitive readout \footnote{The terminology of cavity-based readout of superconducting and semiconducting
qubits is confusing. The literature in both communities uses the term
``dispersive'' readout (because the cavity is pulled in one direction
or another by the state of the qubit) but the physical coupling between
the qubit and the cavity may be completely different.} with Pauli-blockade gives a proven means of differentiating the singlet
versus triplet states of the two dots \footnote{Many qubit options from single spins to double and triple dot encoded
qubits can be read out by conversion to a singlets or triplets of
two dots.}. If you detune the dots into the Pauli-blockade regime (where the
(1,1) state equals the (0,2) state of the double dot system), one
can distinguish singlet from triplet: the singlet state will tend
to allow two electrons to go into the lower dot, when combined the
dots are in the triplet state, they will each stay in their respective
dots as the transition won't be allowed \cite{PettaCoherentManipulationCoupled2005}.
This is sensitive to a readout window given by the temperature and
the singlet-triplet relaxation time, but the signal can still be strong
\cite{Zheng:2019aa}. 

Another option for qubit readout, as advocated by us recently \cite{RuskovQuantumlimitedmeasurementspin2017},
is to attempt quantum curvature readout deep in the (1,1) regime,
where the curvature of the singlet and triplet states is detected
within the S-T relaxation time \cite{ShimChargenoiseinsensitivegateoperations2016,RuskovQuantumlimitedmeasurementspin2017}.
This approach has the benefit of being quantum non-demolition (the
qubit is preserved) and the symmetry of the dot decreases sensitivity
to charge noise and increases S-T relaxation time (because the transition
dipole matrix elements between S and T vanish). It should also be
noted (although this is the first place we have noted it), that this
approach has some immunity to temperature - so may be the most compatible
readout approach for high temperature qubits in relatively small magnetic
fields.

Exchange is the fundamental interaction between electrons related
to the Pauli exclusion principle that allows for fast two-qubit gates
with large ON/OFF ratios. As changing the gate potentials results
in the electron wave functions in the two dots overlapping, the spin
state of the combined system evolves. If this interaction is timed
just right a given two-qubit spin operation can be achieved resulting
in an entangling gate between the qubits \cite{LossQuantumcomputationquantum1998a}.
Quantum operations can be analyzed by measuring Rabi oscillations
when the tunnel barrier is lowered to turn on the exchange interaction.
The latter, if done at the symmetric operating point, allows one to
characterize the charge noise of the device as once you turn on exchange,
the spins are no longer ``protected'' to noise on their wave functions.
More oscillations are indicitave of less charge noise. A useful variant
of this latter experience is to perform a fingerprint plot (see Figure
\ref{fig:multidot_exp}c). 

\section{Caveat Emptor}

The cartoons drawn here are just that. Realization will be more difficult
than imagined and different than proposed. But ideally, the separation
of material optimization from qubit formation will help push the limits
of possible fidelity, or yield, or valley splitting, or whatever is
limited by the material properties. Even when realized and assuming
success, there will still be concerns in translating knowledge gained
from the gate-chip induced dots as compared to the ``real thing.''
Assuming the wafer has been optimized fully, there could be drastically
different results when the dots are fabricated in a more scalable
manner. 

Materials science still matters. Fabricating the gates with the associated
interfaces, dielectrics, processing steps (e.g., anneals) will affect
critical parameters like valley splitting and charge noise. Complex
surface physics due to the passivation of the silicon/silicon-germanium
surface can create unwanted potentials on the quantum dot plane in
the quantum well case or a charge noise environment different than
or worse than a full gate stack. Our hope is that the surface can
be treated in such a way to minimize negative impact.

Shaking of the gate-chip will result in moving of the electrons (with
some symmetry) if the gate-chip is not mounted statically to the dot
wafer. We have not seriously considered the possible implications
of this vibration on dot and quantum operation parameters. Although
any length scales of movement are likely much larger than the dots,
and a much lower relevant frequency compared to gate speed (nanoseconds).
There can also be a vacuum penalty: if the wire is too far from the
surface than the potential of the dot can be washed out. Our point
design simulations indicate that 10 nm separation still allows for
sufficient dot confinement (while STM tunneling typically occurs \textasciitilde 1
nm from the surface distances, \textasciitilde 10 nm can be gauged
with a field emission current). Static mounting, intimate connection
(via a 2D flip-chip instead of perpendicular), and fast readout can
mitigate these concerns.

The approach still requires dilution fridge temperatures unless the
qubits can operate and be operated on at higher temperature (more
below). Qubit operations (e.g., an encoded CNOT gate made up of 20+
pulses) are just as difficult control-wise. Metal shields (that is,
a multi-layer gate chip) above and below the dot gates may be needed
to decrease cross capacitance or improve screening.

A linear array of qubits limits scalability. Introducing longer distance
couplers can connect arrays in a 1.5 dimensional geometry (still linear
but with long-distance coupling via resonators and select dots). To
keep things simple we have discussed one layer of metal on the gate
chip but the concept can get more complicated (wafer to wafer integration
instead of perpendicular chips; multiple layers on the gate chip,
etc). 

\section{What would I do?}

Physics.

1) Let's characterize quantum-relevant wafer properties, especially
valley splitting, charge noise, and disorder, across enough wafers
and ``devices'' to be statistically conclusive.

2) Investigate proposed qubit approaches, encodings, and operation
protocols on one, two, three, and four dot systems \cite{LossQuantumcomputationquantum1998a,DiVincenzo:2000uz,bacon_PhysRevLett.85.1758,FongUniversalQuantumComputation2011,TaylorElectricallyprotectedresonantexchange2013,ShimChargenoiseinsensitivegateoperations2016,RussThreeelectronspinqubits2017,Russ_PhysRevLett.121.177701}.
There are too many unexplored proposals to enumerate, but we still
don't understand in practice which qubit encodings offer the best
trade-offs for qubit quality and classical overhead (number of dots,
pulses, etc), including protocols to minimize leakage \cite{FongUniversalQuantumComputation2011,PhysRevB.91.085419,langrock2020resetifleaked}
and explore measurement-based schemes \cite{brooks2021hybrid}. In
particular, proposals for encoded qubit interconversion and noise-insensitive
always-on, exchange-only qubits \cite{ShimChargenoiseinsensitivegateoperations2016}
could be validated or dismissed. A ten dot device would be sufficient
to implement the vast majority of qubit and gate proposals.

3) Explore alternative readout and coupling approaches via classical
and quantum cavities: transverse versus longitudinal coupling versus
modulated longitudinal coupling \cite{PeterssonChargeSpinState,ReillyFastsinglechargesensing,CollessDispersiveReadoutFewElectron2013,RuskovQuantumlimitedmeasurementspin2017,Ruskov_PhysRevB.103.035301}. 

4) Investigate different materials for their relevance to quantum
computing: \emph{Optimizing valley splitting}: various proposals have
been made to increase valley splitting. Because all of them depend
on the microscopic details of the heterostructure stack, many devices
will need to be measured to have confidence in a solution. \emph{Holes}:
Holes exist at the $\Gamma$-point of the valence band, so there are
no valley splitting issues (although there may be spin-orbit bands
nearby). \emph{Germanium}: germanium has a lower effective mass for
holes (as compared to electrons in silicon) which would relax the
gate wire pitch requirements (and sensitivity to disorder); using
holes in Ge may offer larger spin-orbit coupling as well as no complicating
valley splitting physics. Very recent progress in germanium quantum
dots has been promising \cite{2020-veldhorst-4qubitgermanium,2019-luhman-holedots,2018-gedotsandSC-veldhorst}.
\emph{III-Vs}: Although III-Vs suffer from spinful nuclei, they offer
a benefit of a direct band gap (also plausibly realized in SiGe superlattices
by the way). The gate-chip approach would allow continued cost-realistic
research in III-Vs for optical conversion or for other materials,
such as GaN. \emph{II-VIs}: II-VIs offer the potential for quantum
well dots with spin-0 nuclei and a direct band-gap, they are notoriously
difficult to fabricate. \emph{ZnO} and other oxide-based 2DEGs have
shown inklings of relevance to quantum devices. 2D materials such
as graphene and van der Walls heterostructures (layers of 2D materials)
offer a very large phase space of possible dot implementations (with
such materials, loss may be minimized), the approach here would greatly
accelerate exploration of such materials. \emph{Super-Semi materials:
}In proximitized superconducting-semiconductor stacks, there are opportunities
to explore different approaches to qubit formation via the split-chip
approach presented here while minimizing loss in the substrate. \emph{Topological
materials}: Many potential topological materials are fragile to lithographic
and gate processing. 

5) Study high temperature qubits. Spin qubits continue to have long
coherence times even at elevated temperatures relative to the Zeeman
splitting \cite{PhysRevB.89.075302,PhysRevLett.121.076801,THESIS-Tahan,PetitUniversalquantumlogic2019,2020-yang-dzurak-1kelvinoperation}.
An open question is how robust a 2-qubit gate can be at elevated temperatures
(350 mK or 1-4K). 

6) Search for non-QC applications of these small quantum systems,
such as an for current standards \cite{2012-ritchie-amperedots,2019-NISTampere}.
This approach may allow for far easier fabrication and potentially
better charge noise characteristics.
\begin{figure}
\centering{}\includegraphics[scale=0.25]{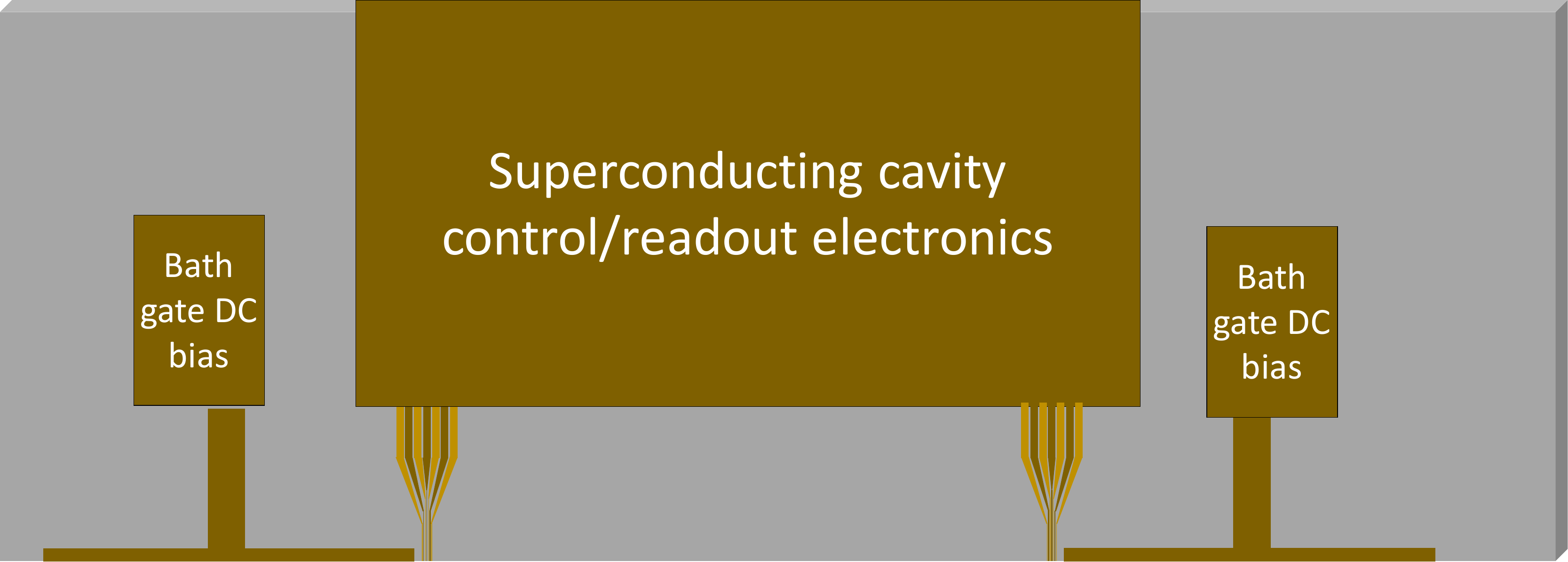}\caption{\label{fig:superchip} \textbf{Superconducting cavity coupling of
2 encoded dot qubits. }It's just as easy to put a superconducting
metal on the gate-chip.}
\end{figure}

\section{One more thing: coupling spins to a superconducting cavity}

We've already discussed using superconducting resonators for readout,
we can go further by exploring spin- qubit entangling protocols \cite{Srinivasa_PhysRevB.94.205421,Ruskov_PhysRevB.103.035301}\textbf{
}via superconducting cavity or transmission lines. By putting the
cavity on the gate-chip, see Figure \ref{fig:superchip}, we can optimize
for high Q. Certain entangling protocols may benefit from high-Q resonators.
Resonators deposited on typical SiGe dot wafers, for example, tend
to have Q's \textless\textless{} 100,000 as compared to millions
achieved on clean sapphire or silicon wafers. A similar approach can
be made with a flip-chip resonator and a traditional dot chip, but
our approach should make fabrication much simpler. It should also
allow networking of small qubit registers enabling a 1.5D quantum
geometry. 

\section{End Speech}

The introduction to virtually every silicon qubit paper goes something
like this: silicon quantum dot spin qubits provide a promising platform
for large-scale quantum computation because of their high-density,
compatibility with conventional CMOS manufacturing, and long coherence
times due to enriched $^{28}$Si material and low spin-orbit coupling.
The future of silicon quantum computing is strong, more-so given recent
progress. Our difficulty has been that we must immediately go to the
final dot dimensions just for the dots to work, we can't push it off.
Qubits need to be small, materials need to be right, and microscopic
effects matter immediately. Many of these problems will eventually
need to be addressed in superconducting qubits, but now you can avoid
them to make progress. 

It's become all the rage to be building ``quantum testbeds''. These
testbeds put as many of the best qubits we have today together in
order to run small algorithms, to achieve quantum supremacy! This
will be exciting, and inconclusive, for some time. Going forward on
this path is obviously necessary, and also toward the first true quantum
error corrected logical qubit. Here I have in mind a different form
of quantum testbed. My testbed can be used to improve or assess new
materials stacks for qubits. It can be used to build few qubit system
to test new designs. It is optimal for a materials-design-test cycle.
Because it separates qubit design (gate structure) from wafer growth,
both can be made better simultaneously. Shifting to a completely different
type of material (e.g. holes in germanium versus electrons in silicon)
requires at most a gate pitch change. There's also no reason one can't
use this approach to make small quantum computers. 

In summary, let us find a way to make and measure more semiconductor
qubits, one way or another.
\begin{acknowledgments}
My highest thanks to my close collaborators Yun-Pil Shim and Rusko
Ruskov whose foundational work is realized in Ref \cite{article-shim-induced-dot-APL}.
Thanks to Hilary Hurst who thought about this idea with me on her
summer visit to LPS in 2016 as part of her National Physical Sciences
Consortium (NPSC) graduate fellowship. Thanks to Bob Butera, Matt
Borselli, Gil Herrera, Bruce Kane, Will Oliver, Ben Palmer, Rusko
Ruskov, and Yun-Pil Shim for critical reading and comments and special
thanks to Tim Sweeney for useful early conversations and motivation.
\end{acknowledgments}

\bibliographystyle{unsrtnat}
\bibliography{spindemo5}

\end{document}